# Real-Time Piano Note Frequency Detection Using FPGA and FFT Core

Shafayet M. Anik and D.G. Perera

ECE Department, University of Colorado Colorado Springs

Colorado Springs, Colorado, USA

## Abstract

Real-time frequency analysis of musical instruments, such as the piano, is an essential feature in areas like electronic tuners, music visualizers, and live sound monitoring. Traditional methods often rely on software-based digital signal processing (DSP), which may introduce latency and require significant computational power. In contrast, hardware platforms such as FPGAs (Field Programmable Gate Arrays) offer the ability to perform such analyses with greater speed and determinism due to their parallel processing capabilities. The primary objective of this project was to analyze analog audio signals from a digital piano using an FPGA-based real-time Fast Fourier Transform (FFT) system.

## Section 1: Introduction

### Context and Problem Overview

Real-time frequency analysis of musical instruments, such as the piano, is an essential feature in areas like electronic tuners, music visualizers, and live sound monitoring. Traditional methods often rely on software-based digital signal processing (DSP), which may introduce latency and require significant computational power. In contrast, hardware platforms such as FPGAs (Field Programmable Gate Arrays) offer the ability to perform such analysis with greater speed and determinism due to their parallel processing capabilities.

For this project, the focus is on identifying the fundamental frequency of a single piano note using a hardware-only solution. A key challenge is efficiently sampling the analog audio signal, performing spectral analysis, and outputting a meaningful result – all within the constraints of a student-accessible FPGA development board.

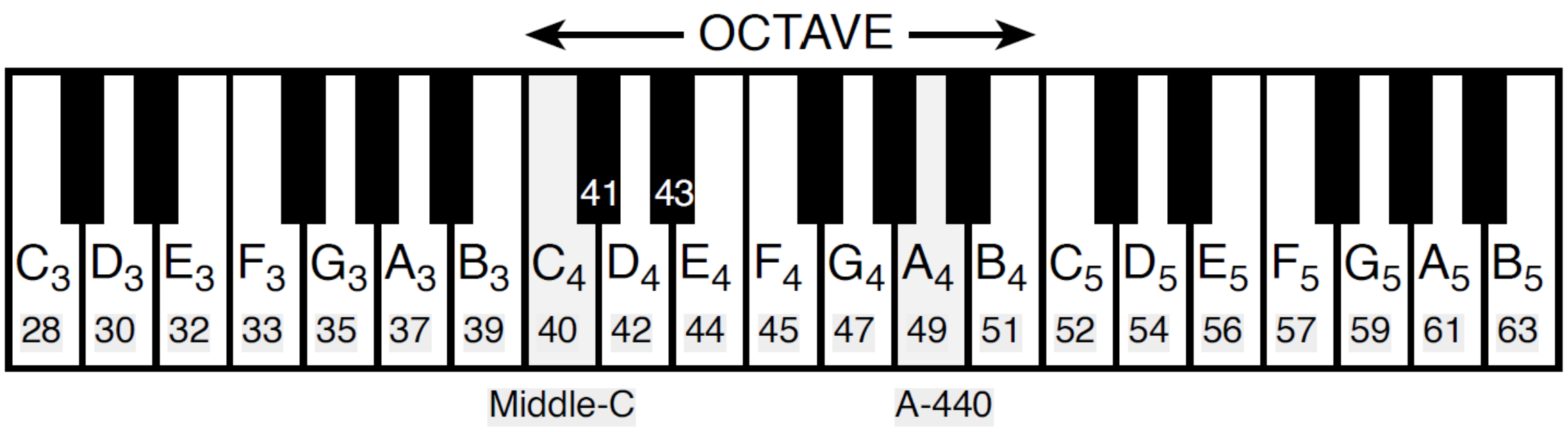

**Figure 1: Layout of a piano keyboard. Key numbers are shaded. The notation C4 means the C-key in the fourth octave.**

### Why This Project Was Chosen

The idea for this project originated from an interest in digital music systems and embedded signal processing. Instead of relying on software tools or high-end processors, we wanted to explore the potential of an FPGA-based solution. The Spartan-3E FPGA board provides a solid platform with built-in support for analog signal capture and digital output, making it a suitable candidate for this kind of implementation.

Another motivation was to gain practical experience working with IP cores, such as the Fast Fourier Transform (FFT), and to manage dataflow between analog input, processing logic, and display output in a hardware-centric design flow.

### Summary of What Was Built

The project implements a piano note detector using an FFT-based approach on an FPGA. The analog audio signal is input through the board's audio capture circuit and preamplifier, then sampled using an ADC module. The captured digital samples are fed into an FFT IP core to identify the dominant frequency component. Once processed, the resulting peak frequency bin is scaled to its corresponding frequency in hertz and displayed on a character LCD screen. The entire process is handled on the FPGA using Verilog modules for each functional block.

**Main Objective**
The overall goal is to build a resource-efficient, real-time hardware system that can capture and process an analog piano tone and display its frequency. The solution should be fast, responsive, and accurate enough to reflect fundamental frequencies commonly produced by acoustic piano keys, while utilizing standard IP blocks and hardware logic.

## Section 2: Project Description, System Diagram, and Design Flow

### Project Overview
The purpose of this project is to design a piano tone analyzer on an FPGA using a modular, hardware-only pipeline. The main function is to capture analog audio input (from a piano or similar tone), convert it into digital form, analyze the frequency content using an FFT core, and display the detected frequency on an LCD. The system was implemented on a Spartan-3E FPGA development board and makes use of several IP cores and custom Verilog modules.

The design takes advantage of the board's built-in analog capture front-end (AFE), which includes a programmable gain amplifier (LTC6912) and a dual-channel ADC (LTC1407A). The FFT core from Xilinx's LogiCORE library is configured to accept 512-point input in burst mode and operate using fixed-point arithmetic. A simple finite state machine coordinates the sampling, FFT processing, and LCD output.

**System-Level Block Diagram**

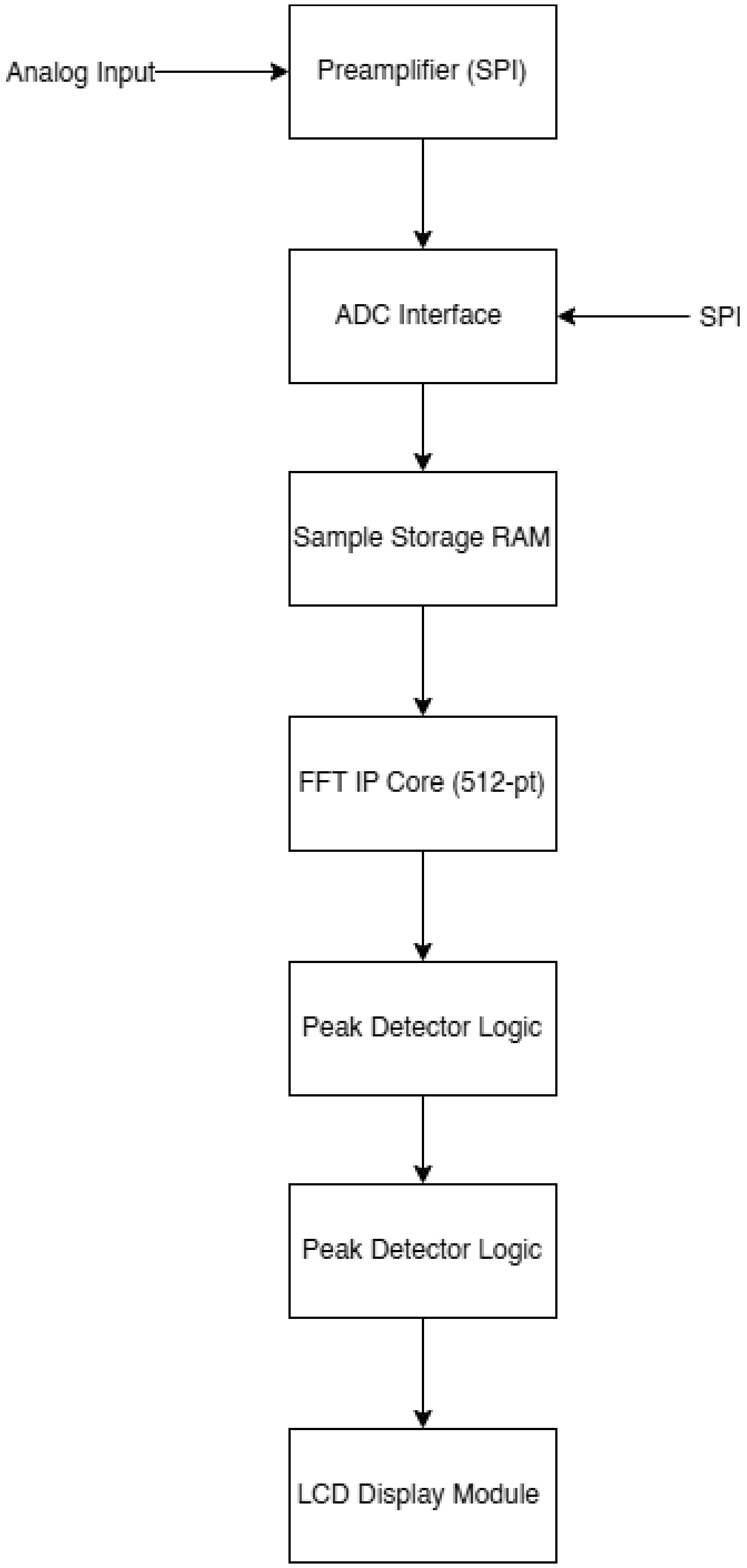

**Module Breakdown**

- **Preamplifier SPI Control**: Configures the gain of the input analog signal using one-shot debounced inputs.

- **ADC Master**: Interfaces with the ADC chip via SPI and collects digital samples at regular intervals.
- **Sample Memory**: A register array stores 512 samples. Sampling is triggered using a debounced pushbutton, and a downsample counter is used to reduce oversampling.
- **FFT Block**: The Xilinx FFT core is configured in Radix-4, Burst I/O mode with 14-bit input and 16-bit phase width. Data is fed in via a loop using the rfd (ready-for-data) signal, and output is received when dv (data valid) is high.
- **Peak Frequency Detector**: A simple max magnitude tracker compares squared magnitudes ($xk_re^2 + xk_im^2$) and stores the highest value and its corresponding index.
- **LCD Module**: The bin index is converted into an estimated frequency (using the FFT resolution in Hz) and passed to a character-based LCD module. It runs as a separate FSM and handles 4-bit command/data sequencing.

**Design Flow**

1. **Initial Planning**: The problem was scoped by listing hardware interfaces and verifying feasibility on the Spartan-3E board. Pin mappings for SPI, ADC, and LCD were reviewed in the user guide.
2. **IP Core Configuration**: The FFT core was customized using Xilinx CoreGen. A testbench was used to verify correct bin placement for simple sinusoidal inputs.
3. **Verilog Design and Simulation**: Each module (ADC master, FFT interface, FSM, etc.) was developed and tested incrementally. Debugging was done using ChipScope ILA and waveform inspection.
4. **Integration and Testing**: All modules were connected in a top-level file. The system was tested using an AUX cable connected to a signal generator app or a keyboard to simulate piano notes.
5. **Output Calibration**: FFT bin indexes were scaled to real frequencies based on the known sampling rate and number of FFT points.
6. **LCD Integration**: After FFT output stabilized, the detected frequency was passed to an FSM that controlled the character LCD to show the detected value.

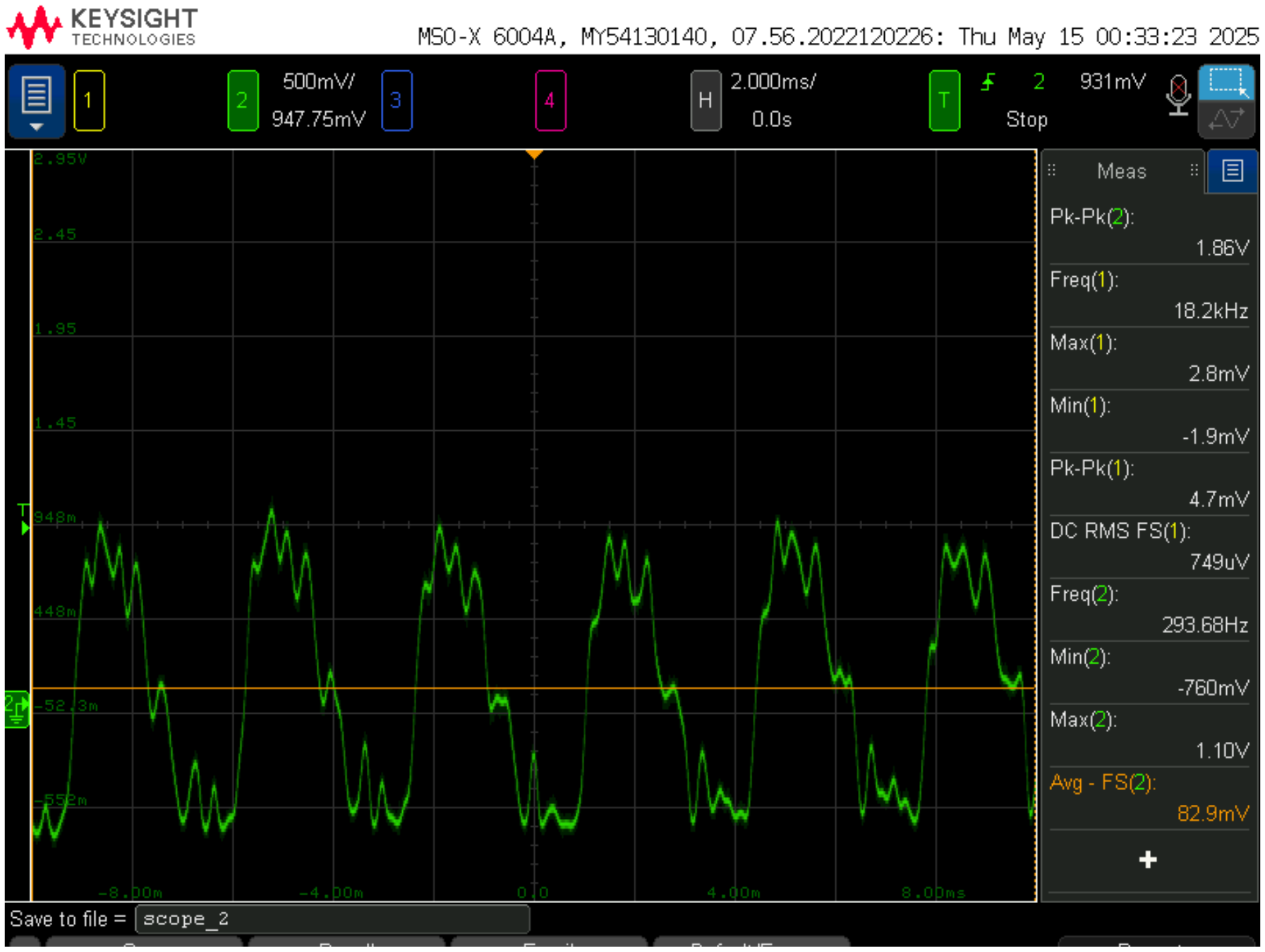

**Figure 2: Waveform Captured for a Piano Key with Detected Frequency at 293.68 Hz**

The oscilloscope shows the time-domain signal for a piano key press, with a fundamental frequency measured at approximately 293.68 Hz.

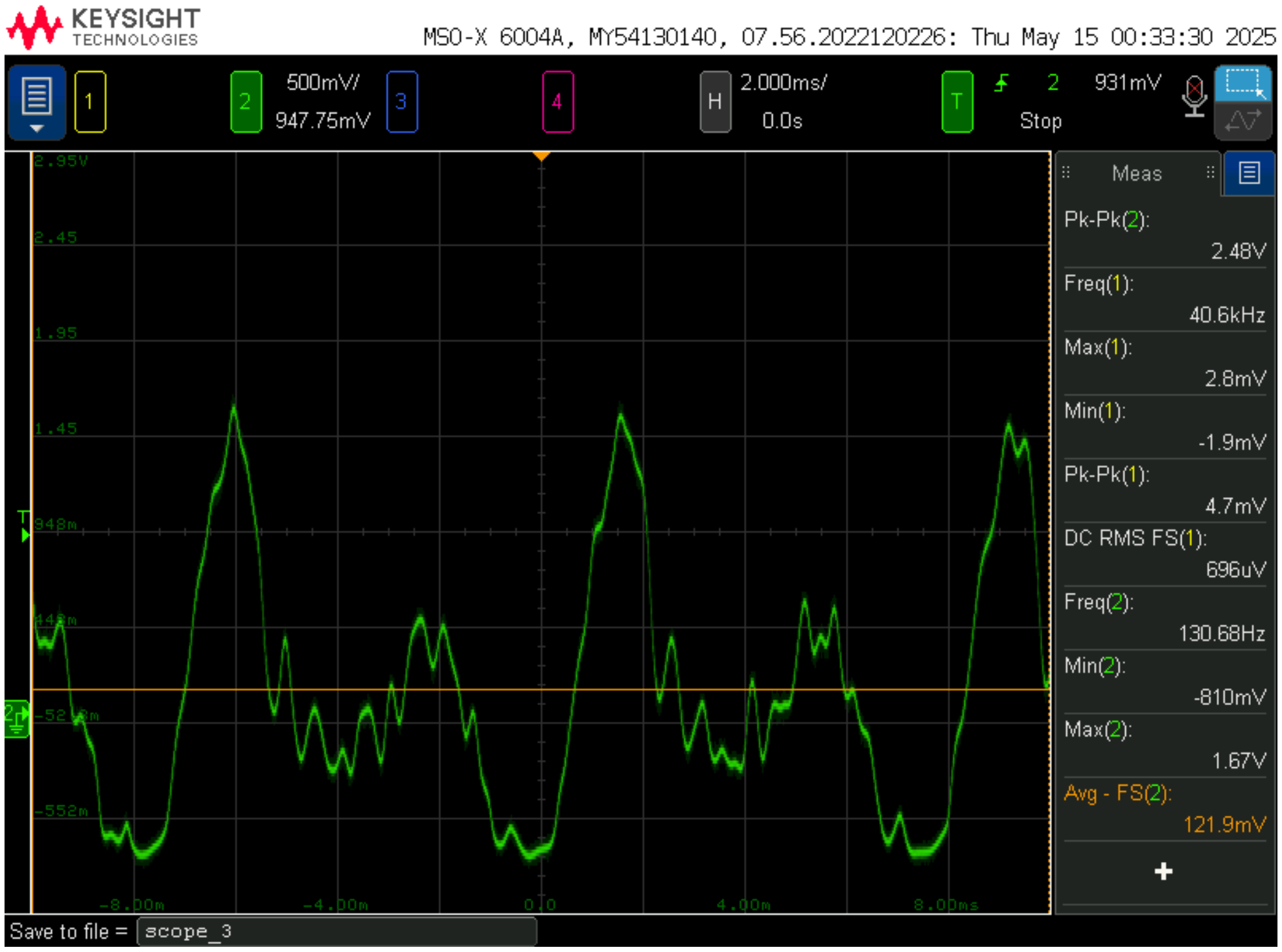

**Figure 3: Waveform Captured for a Different Piano Key with Detected Frequency at 130.68 Hz**

This waveform corresponds to a lower piano note, where the measured fundamental frequency is approximately 130.68 Hz.

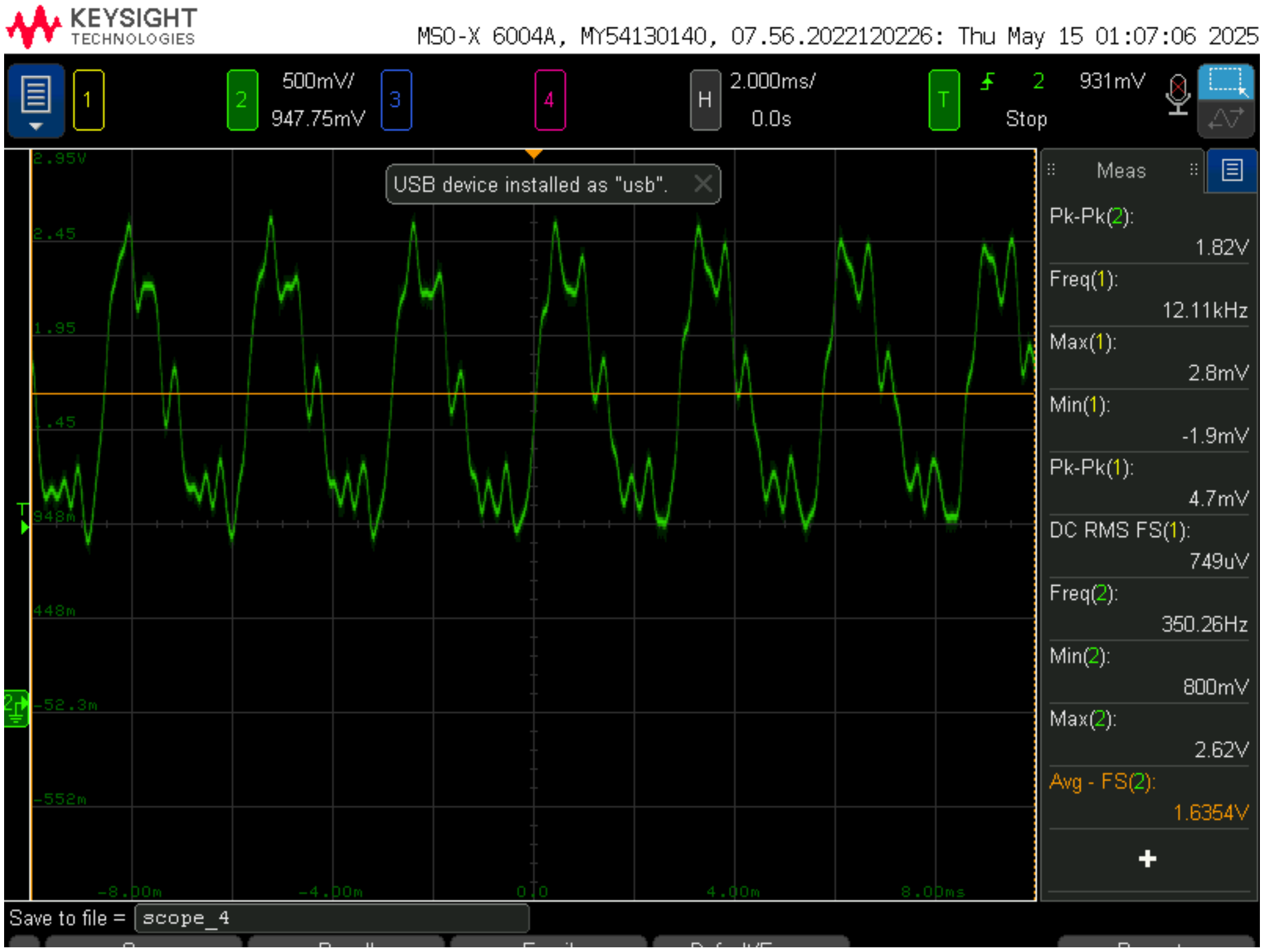

**Figure 4: Waveform Showing 350.26 Hz Frequency with DC Offset Bias of 1.65 V for ADC Compatibility**

The waveform includes a DC offset of 1.65 V intentionally added by the signal generator. This is required because the Spartan-3E FPGA's ADC interface expects input signals within a unipolar range centered around the internal DC voltage reference (approximately 1.65 V). The DC bias ensures the full swing of the sinusoidal input is captured correctly without clipping or distortion.

## Section 3: Hardware Design and Implementation

The hardware design for this project was centered around the **Spartan-3E FPGA Starter Kit**, which offered sufficient resources for signal capture, digital processing, and output display. The main goal of the hardware design was to implement a real-time frequency detection system that could take analog signals from a musical instrument (like a piano), convert them into digital samples using the onboard ADC, process them using a Fast Fourier Transform (FFT) block, and display the peak frequency using an LCD.

---

### System Overview

The hardware system was divided into the following main modules:

- **Pre-Amplifier and ADC Interface**
- **SPI Data Acquisition**
- **FFT Engine (Xilinx Core)**
- **Peak Frequency Extraction**
- **LCD Display Output**

Each module was developed incrementally in Verilog HDL and tested individually using simulation and ChipScope.

---

### Module Descriptions

#### 1. Pre-Amplifier and ADC Interface

The Spartan-3E board includes a built-in Linear Technology LTC6912 programmable gain amplifier and LTC1407A ADC. The analog input was taken from the piano's audio output using an AUX cable. To ensure compatibility with the FPGA's input range (centered around 1.65V), a DC offset was manually added using a waveform generator.

The ADC output was captured over an SPI interface, and only Channel A was used. The sampled values were stored in a buffer inside the FPGA.

#### 2. SPI Master (ADC + PreAmp Control)

SPI communication was handled using a custom Verilog SPI master module. The pre-amp gain setting and ADC sampling trigger were controlled through toggling SPI signals. Timing issues were carefully handled by inserting debounced one-shot triggers before each SPI command.

#### 3. FFT Engine

#### Theory of Operation

The FFT is a computationally efficient algorithm for computing a Discrete Fourier Transform (DFT) of sample sizes that are a positive integer power of 2. The DFT X(k), k = 0, …, N − 1 of a sequence x(n), n = 0, …, N − 1 is defined as:

$$X(k) = \sum_{n=0}^{N-1} x(n) e^{-jnk2\pi/N} \quad k = 0, \ldots\ldots, N-1$$

(Equation 1)

where N is the transform size and j = √−1. The inverse DFT (IDFT) is given by:

$$X(k) = \frac{1}{N} \sum_{n=0}^{N-1} x(n) e^{jnk2\pi/N} \quad k = 0, \ldots\ldots, N-1$$

(Equation 2)

**Algorithm**

The FFT core uses the Radix-4 and Radix-2 decompositions for computing the DFT. For Burst I/O architectures, the decimation-in-time (DIT) method is used, while the decimation-in-frequency (DIF) method is used for the Pipelined, Streaming I/O architecture. When using Radix-4 decomposition, the *N*-point FFT consists of log4 *(N)* stages, with each stage containing *N/4* Radix-4 butterflies. Point sizes that are not a power of 4 need an extra Radix-2 stage for combining data. An *N*-point FFT using Radix-2 decomposition has log*2* (*N)* stages, with each stage containing *N/2* Radix-2 butterflies.

**Configuration**

The Xilinx FFT IP core was configured in **Radix-4, Burst I/O** mode with a **512-point transform**. The input width was set to 14 bits and the phase factor width to 16 bits. Initially, we tried using a 1024-point FFT, but resource usage exceeded available slices on the Spartan-3E, so it was scaled back to 512.

The FFT block required feeding one sample per valid RFD cycle and generating results during DV high. Unload logic was included to align with the burst input mode.

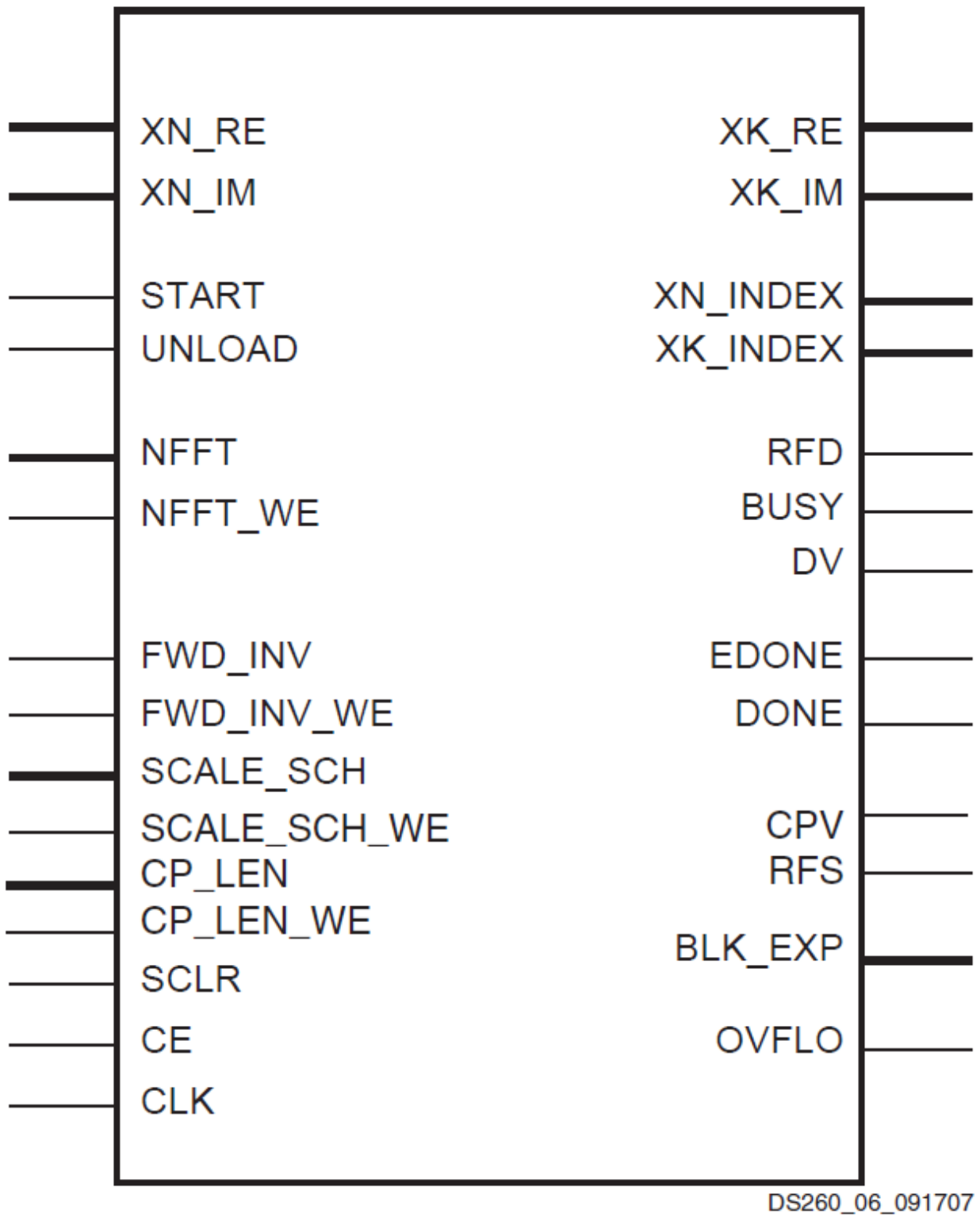

**Figure- 5 : Core Schematic Symbol (Single Channel)**

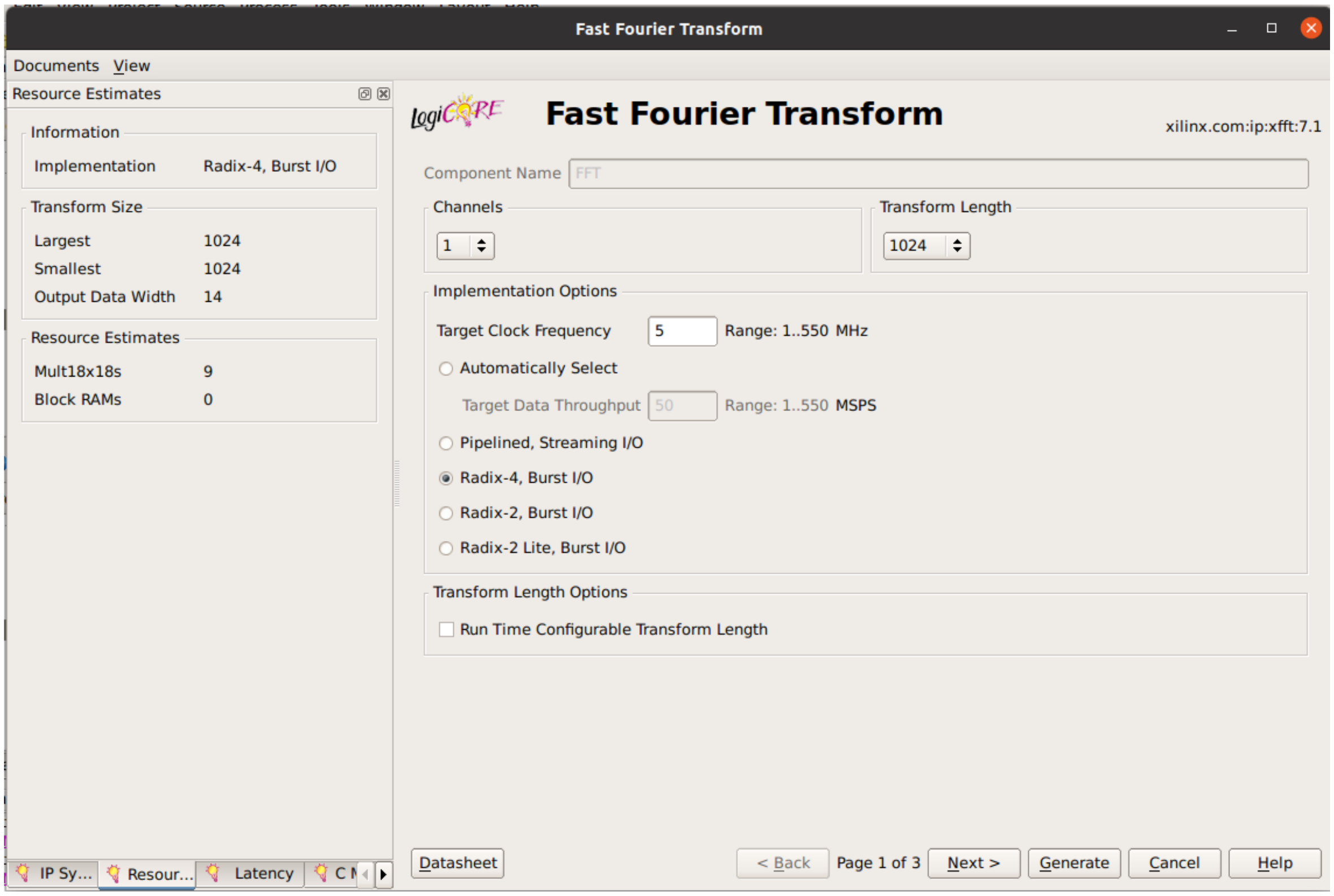

**Figure 6: Configuration of FFT Core with 1024-Point Transform in Xilinx Core Generator**

This figure illustrates the setup of the FFT IP core, showing parameters such as transform length, implementation mode (Radix-4 Burst I/O), and clock frequency (5 MHz). The transform size was initially set to 1024 points, with a fixed-point 14-bit output format.

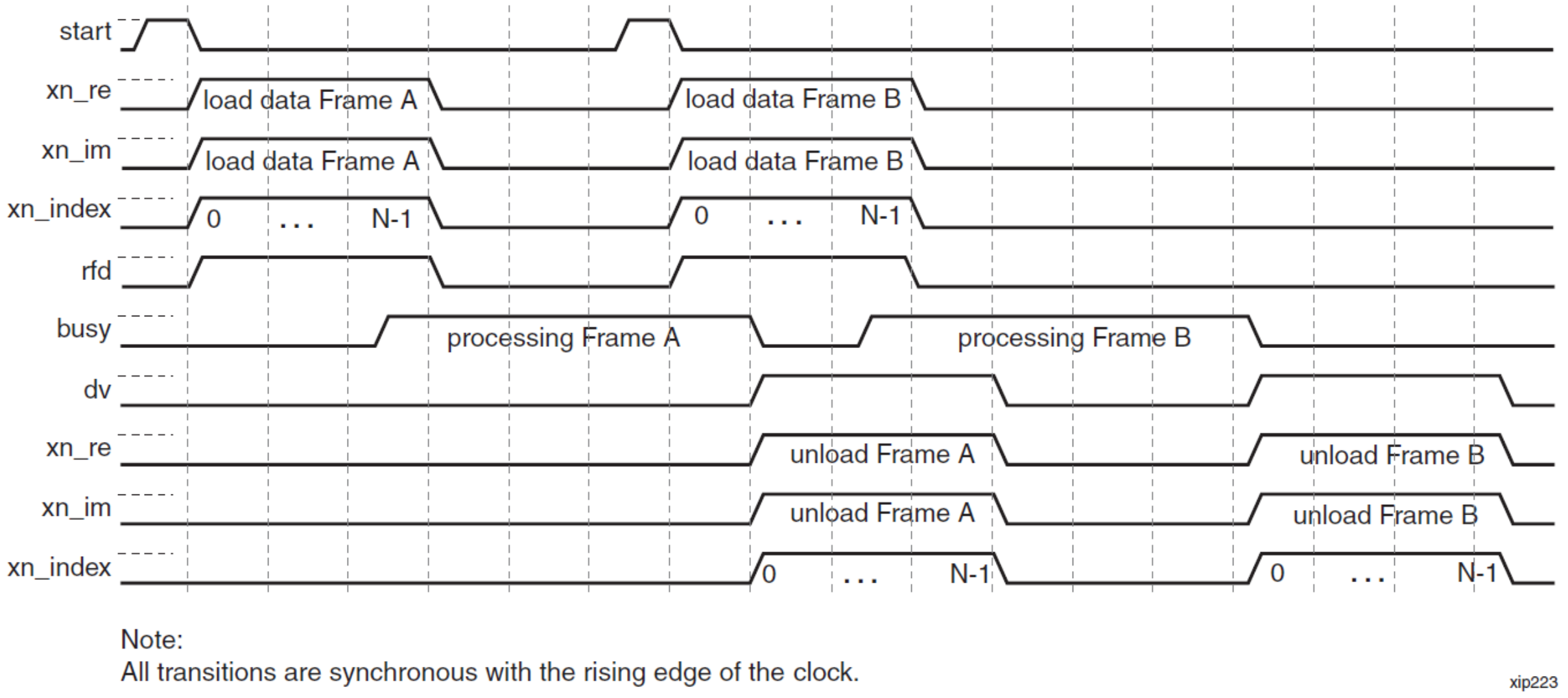

**Figure-7: Timing for Non-Continuous Data Stream**

This waveform diagram illustrates the timing behavior of the Xilinx FFT IP core operating in Burst I/O mode, specifically during two successive data frames (Frame A and Frame B). It captures how control and data signals behave over time relative to the system clock.

**Explanation of Key Signals:**

- **start:** This signal initiates the FFT operation. A rising edge indicates that the input frame (e.g., Frame A or B) is ready to be loaded. It must be asserted for at least one clock cycle.
- **xn_re and xn_im:** These are the real and imaginary components of the input data stream. When rfd (ready for data) is high, valid input samples are presented on these buses.
- **xn_index:** Indicates the index of the input data being supplied to the FFT core. It starts from 0 and increments to N-1 (where N is the transform size) during the data loading phase.
- **rfd (Ready for Data):** When high, the core is ready to accept input data. It stays high during the data load period and goes low once the frame is fully loaded.
- **busy:** This signal goes high to indicate that the core is processing the FFT internally. No new data is accepted during this time.

- **dv (Data Valid):** This signal indicates that output data is now available on the xk_re/xk_im output buses. It is asserted once the internal processing of the frame is complete and stays high during the unloading of results.
- **xk_re and xk_im:** Real and imaginary parts of the FFT output.
- **xk_index:** Output index, corresponding to the FFT bin for the current output sample.

**Behavior Summary:**

- When **start** is asserted, the FFT core begins accepting data (**xn_re, xn_im**) while **rfd** is high.
- After all N samples are loaded (**xn_index** runs from **0** to **N-1**), **rfd** goes low and **busy** goes high, indicating that the core is performing the FFT computation.
- Once **busy** deasserts, **dv** becomes high and the output data is streamed out sequentially (xk_index from 0 to N-1).
- This cycle repeats for subsequent frames.

**Practical Use:**

This figure is crucial for developers designing state machines to interface with the FFT core. It shows how to synchronize input sampling, processing, and output capturing stages. In this project, similar signal behavior was validated using ChipScope Pro/ILA debugging.

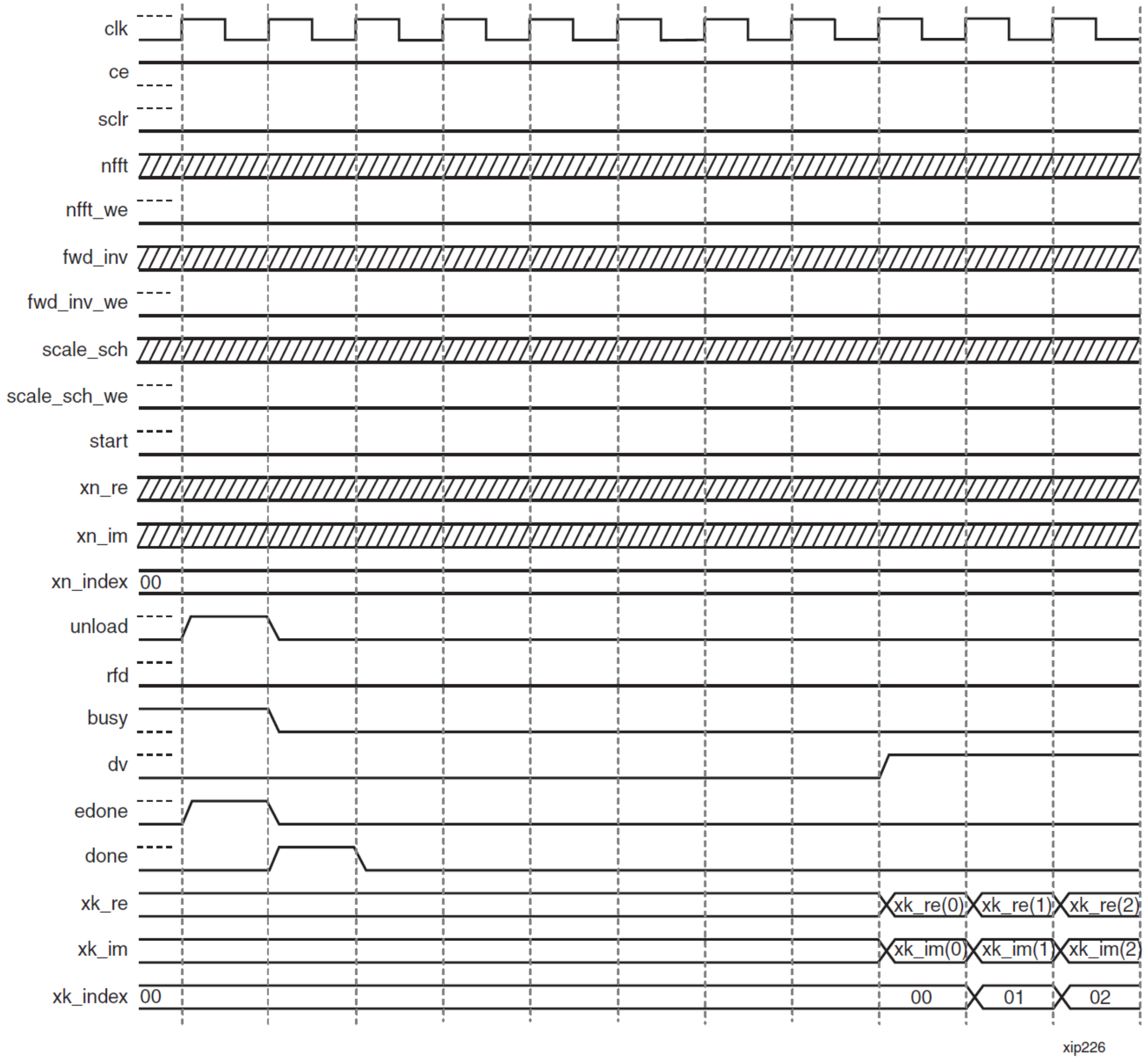

**Figure-8: Unload Output Results in Natural Order**

This timing diagram (Figure 8) illustrates how the **Xilinx FFT core unloads FFT output data in natural order**. The behavior of key control and data signals during the output (unload) phase is shown after the FFT computation is complete.

**Key Signals Explained:**

- **clk**: The main system clock driving the FFT core.
- **ce / sclr**: Clock enable and synchronous clear. These are held inactive in this example.
- **nfft / nfft_we, fwd_inv / fwd_inv_we, scale_sch / scale_sch_we**: Configuration signals for transform length, transform direction, and scaling. These are typically set before asserting start.

- **start**: Triggers the start of FFT data loading and computation.
- **xn_re / xn_im**: Input real and imaginary data buses (not changing here since loading is done).
- **xn_index**: Index counter for input data—stays static after data is loaded.
- **unload**: Goes high to indicate the beginning of output unloading.
- **rfd (Ready for Data)**: Goes low during computation and remains low during unloading.
- **busy**: High during the FFT computation process.
- **dv (Data Valid)**: Goes high when output data is valid and can be read.
- **edone / done**: Indicate the end of data processing and unloading.
- **xk_re / xk_im**: Real and imaginary parts of the FFT result, which begin appearing when dv is high.
- **xk_index**: The index of the current FFT output sample, here showing natural order: 0, 1, 2, ...

**Summary:**

Once the FFT computation is complete (busy goes low and edone pulses), the **unload** signal is asserted to start the data readout. The output data (xk_re, xk_im) becomes valid when dv is high, and the xk_index increments naturally. This mode is ideal when post-processing logic expects data in sequential bin order.

This timing behavior is important to implement correct data capture logic in your design and was replicated in this project using the Burst I/O mode of the FFT core.

**4. Peak Detection**

The peak FFT bin was tracked by squaring the real and imaginary outputs (xk_re, xk_im) and comparing the magnitude with the previously stored peak. If a new higher value was found, the index was stored. This index was then used to approximate the dominant frequency.

**5. LCD Display**

The frequency output was shown on a standard 16x2 LCD using 4-bit mode. A custom LCD controller module was created to convert the 9-bit FFT index into characters. Each update cycle was triggered by a lcd_start signal.

---

**Challenges Encountered**

- **High Resource Usage**
  When using a **1024-point FFT**, the synthesis process failed due to excessive slice utilization—exceeding 100% of available logic resources on the Spartan-3E FPGA. An attempt was also made to configure the FFT core using the **Pipelined, Streaming I/O** mode for faster throughput. However, this configuration required substantial **Block RAM (BRAM)** for continuous data streaming and buffering, which exceeded the available memory on the Spartan-3E board. In addition to BRAM limits, the configuration also consumed a high number of logic slices and DSP blocks. To address this, the design was reverted to a **Radix-4, Burst I/O** implementation and the transform length was reduced to **512 points**, which successfully synthesized within timing and resource constraints.

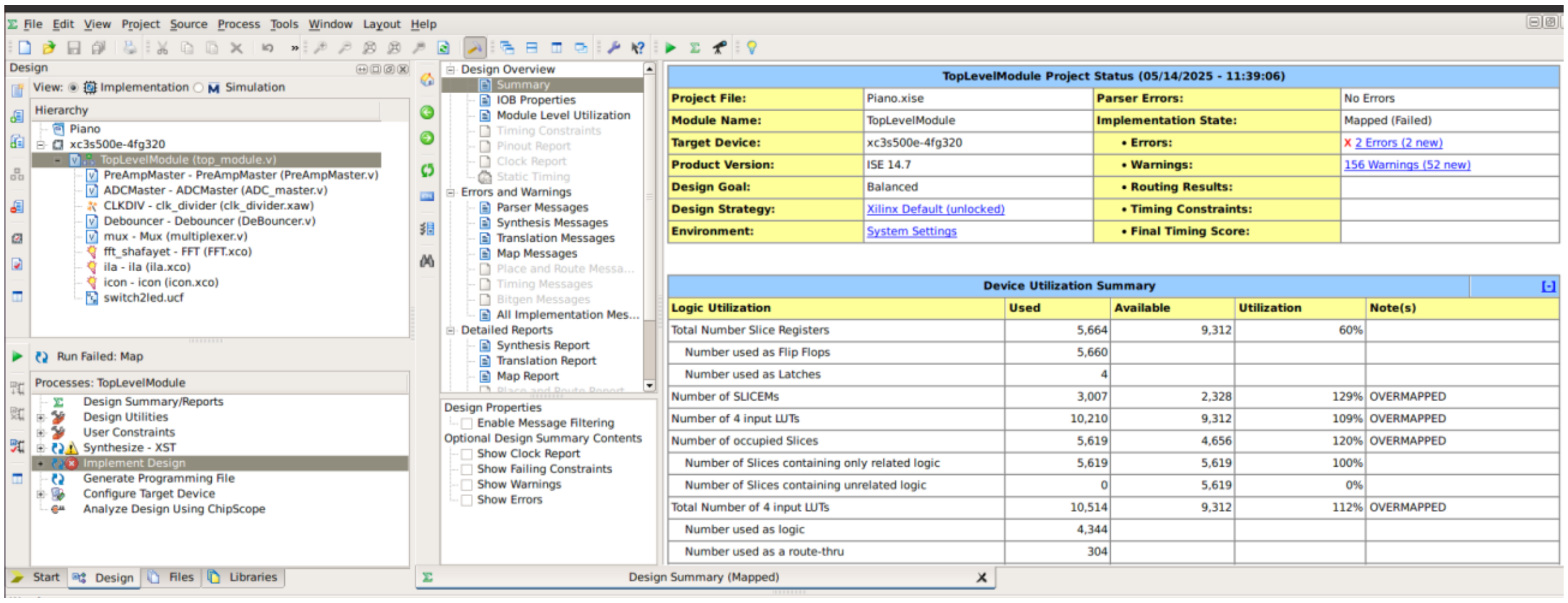

**Figure 9: Post-Synthesis Resource Utilization Showing Overmapped Slices for 1024-Point FFT**

This figure displays the device utilization summary in Xilinx ISE. Due to the high transform length (1024 points), the design exceeded available slices and LUTs, resulting in overmapping and synthesis failure.

- **Noise in ADC Sampling**
  The piano signal was initially very weak. Without proper grounding via the AUX cable sleeve, the signal was distorted. Connecting the ground properly and applying a DC offset resolved the issue.

- **LCD Flickering**
  Initially, the LCD was displaying garbled or flickering characters. This was due to insufficient delays between command instructions. Adding delay states in the FSM fixed this issue.

- **Timing Conflicts**
  Synchronizing the ADC sampling and FFT feeding was tricky due to different module latencies. To solve this, we added a state machine to manage transitions and ensure data was only fed to the FFT when rfd was high.

**Peripherals Used**

- **Keysight Oscilloscope**: To validate the waveform being fed into the ADC.
- **Waveform Generator**: For adding DC offset.
- **LCD Module**: For real-time display of frequency output.

**Section 4: Sampling Rate and Frequency Resolution**

To manage resource constraints and ensure compatibility with the FFT core, the sampling rate was intentionally reduced by inserting delays between ADC captures. Each ADC sample was spaced by **73 clock cycles**, and from those, **only every 16th ADC sample** was stored into memory for FFT analysis. With a system clock of **5 MHz**, this leads to an effective sampling rate of:

$$Fs = \frac{5MHz}{73 \times 16} = \frac{5 \times 10^6}{1168} \approx 4.28KHz$$

Given a **512-point FFT**, the frequency resolution is:

$$\Delta f = \frac{fs}{512} = \frac{4280}{512} \approx 8.36 \text{ Hz/bin}$$

**Section 5: Discussion and Conclusion**

The primary objective of this project was to analyze analog audio signals from a digital piano using an FPGA-based real-time Fast Fourier Transform (FFT) system. Through a combination of hardware modules—including ADC interfacing, SPI communication, FFT IP core, and LCD display logic—the system was able to capture, process, and visualize the frequency domain representation of played notes. Each key press translated into a distinct frequency bin output, confirming correct implementation of the FFT and sampling pipeline.

One important observation was that the frequency resolution heavily depends on both the number of FFT points and the effective sampling rate. During testing, a resolution of approximately 4.28 Hz per FFT bin was achieved using a 512-point FFT with a sampling rate of roughly 4.28 KHz. This proved sufficient for distinguishing common musical notes but left room for optimization, especially when detecting harmonics or notes in close proximity.

A significant design constraint encountered was the limited resources of the Spartan-3E FPGA, particularly the availability of Block RAMs. While the Xilinx FFT IP core supports high-throughput streaming and pipelined architectures, attempting to use these configurations exceeded the available logic and BRAM, resulting in synthesis and map errors. As a result, a more conservative Radix-4, Burst I/O configuration was selected, which offered a compromise between accuracy and resource usage.

**Conclusion:**

The final design met the core functional requirements: sampling analog input, performing FFT, and displaying the dominant frequency on an LCD. It demonstrated that even with resource-constrained FPGAs, real-time frequency domain analysis is feasible using fixed-point FFT and thoughtful optimization of sampling and processing stages.

This work is inspired by the digital design research group at UCCS. This group has done extensive work in FPGA-based architectures, techniques, and associated models. Their analyses [3],[4] shows that FPGA-based systems are currently the best option to support applications and algorithms, such as the ones presented in this report. Also, their previous work on FPGA-based accelerators, architectures, and techniques for various compute and data-intensive applications, including data analytics/mining [5],[6]; control systems [7],[8]; cybersecurity [9],[10]; machine learning [11],[12]; communications [13],[14]; edge computing [15],[16]; bioinformatics [17]; and neuromorphic computing [18],[19]; demonstrated that FPGA-based systems are the best avenue to support and accelerate complex algorithms.

**Future Work:**

For future iterations, the following improvements are recommended:

- **Enhanced Sampling Strategy:** Incorporate adaptive sampling or a dynamic gain controller to handle broader input amplitudes and improve SNR.
- **Streaming FFT Architecture:** If ported to a more capable FPGA (e.g., Artix-7 or Zynq), utilize the Pipelined Streaming I/O version of FFT to support real-time continuous signal analysis.
- **Peak Interpolation:** Implement peak interpolation or zero-padding to achieve sub-bin frequency resolution and improve note accuracy.
- **Multi-note Detection:** Extend the FFT output processing logic to detect and label chords or multiple simultaneous frequencies.

- **Graphical Display:** Replace the character-based LCD with a graphical display (e.g., VGA or OLED) for real-time spectrum visualization.

Also as future work, we are planning to investigate hardware optimization techniques, such as parallel processing architectures (similar to [20],[21]), partial and dynamic reconfiguration traits (as stated in [22]) and architectures (similar to [23]), HDL code optimization techniques (as stated in [24]),  and multi-ported memory architectures (similar to [25]), to further enhance the performance metrics of FPGA-based architectures, while considering the associated tradeoffs.

## Appendix: Project Snapshots and ILA Waveforms

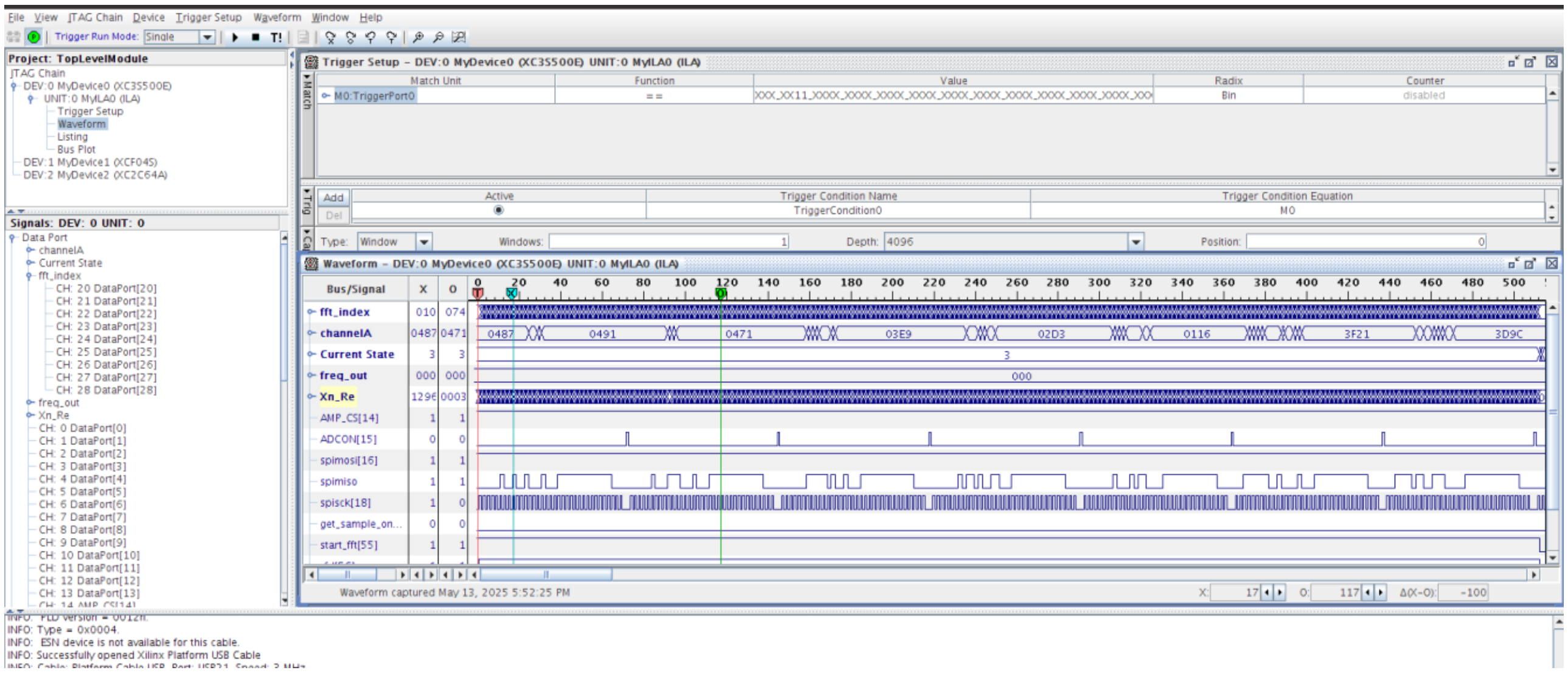

**Figure 10: ILA Debugging Using ChipScope—Verifying ADC Sampling and Signal Activity**

This waveform shows the real-time behavior of control and data lines using ChipScope. It was used to verify correct data capture from the ADC, timing of control signals like get_sample_oneshot, and to confirm transitions between FSM states.

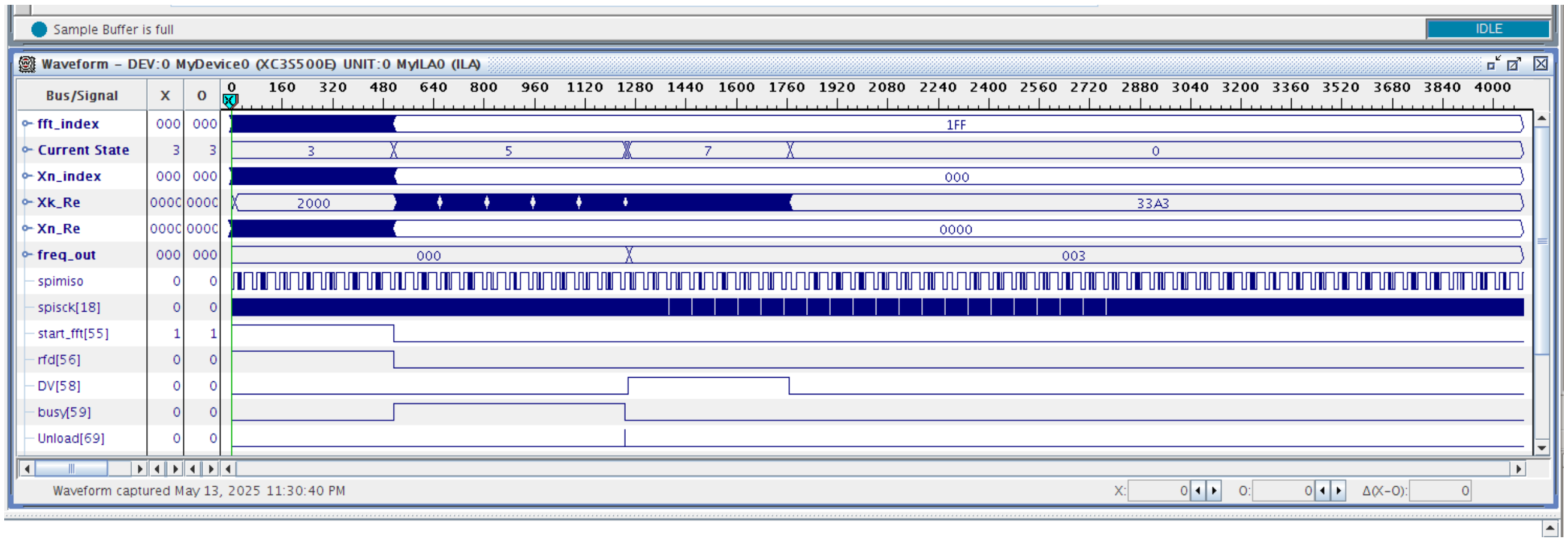

**Figure 11: ChipScope Debugging During FFT Feeding and Processing Stage**

This waveform captures the behavior during FFT processing. It shows signals such as fft_index, xn_index, xk_re, and dv toggling during the feed and unload phases of FFT. The captured data confirms the correct sequencing of FFT input and output.